\documentclass[aps,prl,twocolumn,groupedaddress]{revtex4-1}

\usepackage{bm}
\usepackage{graphicx}
\usepackage{amsmath}
\usepackage{amsfonts}
\usepackage{amssymb}

\begin{document}


\title{Ferrimagnetism of dilute Ising antiferromagnets}

\author{P. N. Timonin}
\email[]{pntim@live.ru}
\affiliation{Southern Federal University,
344090, Rostov-on-Don, Russia}


\date{\today}

\begin{abstract}
It is shown that nearest-neighbor antiferromagnetic interactions of identical Ising spins on imbalanced bipartite lattice and imbalanced bipartite hierarchical fractal result in ferrimagnetic order instead of antiferromagnetic one. On some crystal lattices dilute Ising antiferromagnets may also become  ferrimagnets due to the imbalanced nature of the magnetic percolation cluster when it coexists with the percolation cluster of vacancies. As evidenced by the existing experiments on $Fe_pZn_{1-p}F_2$, such ferrimagnetism is inherent property of bcc lattice so thermodynamics of these compounds at low $p$ can be similar to that of antiferromagnet on imbalanced hierarchical fractal.  
\end{abstract}


\maketitle

The system of the identical Ising spins on the sites of some crystalline lattices with the nearest-neighbor antiferromagnetic (AF) exchange may have magnetized ground states. In such states there would be antiparallel neighboring spins, as interaction dictates, but the whole numbers of up-spins and down-spins would differ.  One such $2d$ lattice is shown in Fig. 1(a). Here two sublattices with parallel up and down spins in the ground state are shown by filled and empty circles correspondingly. We see that in the unit cell there are one filled circle and two empty ones so we get $ \pm 1/3$ magnetizations in two globally-reversed ground states for nearest-neighbor AF on this lattice. 
Thus this AF model has a couple of ferrimagnetic ground states with both staggered $L = \left( {\left\langle {{S_A}} \right\rangle  - \left\langle {{S_B}} \right\rangle } \right)/2$ and homogeneous $M = \left( {2\left\langle {{S_A}} \right\rangle  + \left\langle {{S_B}} \right\rangle } \right)/3$ magnetizations. 

One can easily show that this ordering persists up to finite $ T_c $. Summing the Gibbs function over spins on sublattice A (empty circles) we get the Gibbs distribution for the spins on the sublattice B having effective ferromagnetic Hamiltonian. Indeed, for each link with $ S_A $ spin we have ($J$ is AF exchange)
\begin{eqnarray*}
\sum\limits_{{S_A} =  \pm 1} {\exp \left[ { - {S_A}\left( {{S_B} + {{S'}_B}}{J}/{T} \right)} \right]}  = 2\cosh \left[ {\left( {{S_B} + {{S'}_B}}{J}/{T} \right)} \right] \\
= 2\exp \left[ {{K_B}\left( {{S_B}{{S'}_B} + 1} \right)} \right], 
\qquad 2{K_B} \equiv \ln \cosh \left( {2{J}/{T}} \right)
\end{eqnarray*}
Hence, the ordering of $ S_B $ spins is described by the ferromagnetic Ising model on the square lattice, so $\left\langle {{S_B}} \right\rangle  \ne 0$ for 
${K_B} > \frac{1}{2}\ln \left( {\sqrt 2  + 1} \right)$ \cite{1} or 
\[
T < {T_c} = 2J/\ln \left( {\sqrt 2  + 1 + \sqrt {2\left( {\sqrt 2  + 1} \right)} } \right).
\]
\begin{figure}[htp]
\centering
\includegraphics[height=3.3cm]{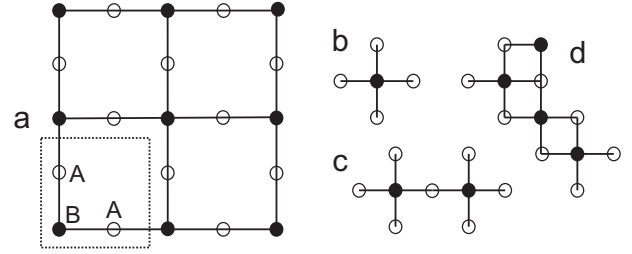}
\caption{ Examples of imbalanced bipartite graphs with different numbers of sites in sublattices A (open circles) and B (filled circles), $N_A > N_B$. (a) - fragment of regular $2d$ lattice, dotted line shows the unit cell; (b, c, d) - clusters of dilute square lattice. In the ground state short-range Ising AF on them would have parallel spins on A and B sublattices and, hence, a nonzero magnetization. }\label{Fig.1}
\end{figure}
As $\left\langle {{S_B}} \right\rangle $ is a linear combination of $L$ and $M$, the ordered phase is a ferrimagnetic one. This implies that homogeneous magnetic field $H$ has a part conjugated with the order parameter so magnetic susceptibility diverges at $T_c$ when $H=0$ and the transition becomes smeared at finite $H$.

Thus we have a simple example showing that nearest-neighbor AF interaction of identical Ising spins may result in the macroscopic ferrimagnetic order. This is in apparent distinction with conventional ferrimagnets having several different magnetic moments in a cell. It may look rather exotic in the realm of real crystals yet such situation can be frequent in disordered Ising AF, first of all, in dilute Ising AF on bipartite lattices. These lattices can be divided in two subsets of sites, A and B, such that all bonds are of the A-B type, i. e. there are no bonds inside A and B subsets \cite{2}. Apparently, the Ising AF on such lattice is non-frustrated having all spins up on sublattice A and down on sublattice B or vice versa in its two degenerate ground states. Their magnetizations are
\[
m =  \pm \frac{{{N_A} - {N_B}}}{{{N_A} + {N_B}}} =  \pm \frac{{1 - \eta }}{{1 + \eta }},{\text{    }}\eta  \equiv \frac{{{N_B}}}{{{N_A}}} < 1
\]
Here $ N_A $ and $ N_B $ are the numbers of sites in A and B sublattices and we choose $\eta<1$ for definiteness. 

Seemingly, all known non-frustrated AF crystals with just one sort of magnetic ions have bipartite lattices that are the balanced ones, that is with $\eta=1$ and purely AF ground states, while Fig. 1(a) shows the imbalanced bipartite lattice with $\eta<1$ ($\eta=0.5$). Yet the dilution of balanced bipartite lattices results in appearance of a number of isolated clusters, mostly with $\eta < 1$, those with $\eta=1$ being the rare exceptions. Figs. 1(b, c, d) show the imbalanced clusters on the square lattice. So at $T=0$ and arbitrarily small magnetic field dilute bipartite AF must show nonzero magnetization due to the presence of such imbalanced finite clusters. This circumstance was first noticed by Neel \cite{3}. Still it stays unnoticed that for some concentrations of magnetic ions $p$ the giant percolation cluster may also have the average imbalance ratio $\eta_p < 1$.

Indeed, in finite sample the role of percolation cluster belong to that with the largest number of sites and very probably it is imbalanced, as most of them. However, in the thermodynamic limit ($N \to \infty$) $\eta_p$ will tend to unity if there are only finite clusters of vacancies. Apparently, such  finite clusters cannot make infinite lattice imbalanced as for every cluster deleting unequal number of sites from A and B sublattices there exists (with the same probability) the shifted cluster of the same form which restores the balance. Thus at $1- p_c < p < 1$ the imbalanced percolation cluster can only exist as a finite-size effect. Meanwhile, at $p_c < p < 1- p_c$ there is infinite percolation cluster of vacancies to which this argument does not apply. Hence, here $\eta_p < 1$  may also hold in the $N \to \infty$  limit in some crystal lattices. Then the ground state magnetization of dilute AF in this interval will be
\[
{m_p}\left( {H =  + 0} \right) = \int\limits_0^1 {\frac{{1 - \eta }}{{1 + \eta }}} {W_p}\left( \eta  \right)d\eta  + \frac{{1 - {\eta _p}}}{{1 + {\eta _p}}}
\]
Here ${W_p}\left( \eta  \right)$ is the imbalance distribution function of finite clusters.
Now it seems that neither ${W_p}\left( \eta  \right)$ nor $\eta _p$ are known for the crystal lattices. So to find them is quite relevant task for the physics of dilute short-range AF. 

The magnetization of finite clusters vanishes at finite temperatures, but that of the imbalanced percolation cluster would persist up to a finite $T_c$ and $M \sim L$ at all $T$ due to its geometrical origin.  Then macroscopic features of dilute AF would be those of ordinary ferromagnet in spite of the presence of antiparallel neighboring spins in the ordered phase. In such a case, the mapping of this model onto random-field Ising magnet (RFIM) \cite{4} would be no longer valid as it suggests purely AF transition in DAFF. This possibility of AF order breaking is missed in Ref. \cite{4} which is a consequence of the mean-field treatment of homogeneous magnetization.  

The evidences in favor of DAFF ferrimagnetism can be found in experiments on several dilute Ising AF with $p < 1 - p_c$ showing the remanent magnetization with the usual order-parameter behavior \cite{5}-\cite{9} and prominent peak in temperature dependence of magnetic susceptibility which appears in low fields as a result of dilution and becomes gradually smeared in higher fields \cite{8},\cite{9}. 

\begin{figure}[htp]
\centering
\includegraphics[height=10cm]{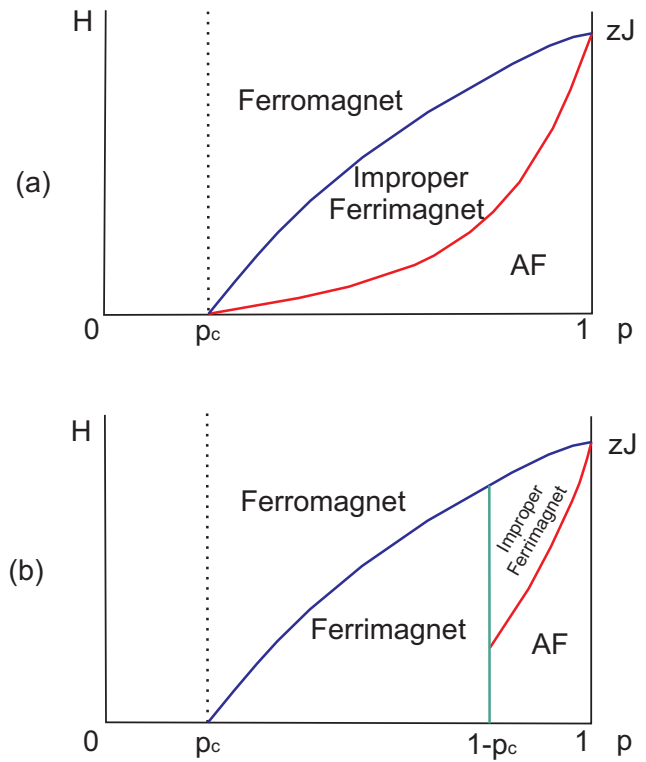}
\caption{Schematic ground state $H-p$ phase diagrams of dilute AF. (a) $\eta_p = 1$ for all $p$, (b) $\eta_p <1$ for $p < 1-p_c$. The lines between phases are defined by ${H_{AF}}( p)$ and ${H_{F}}( p)$ discussed in text. In the improper ferrimagnetic regions sharp AF transition is preserved at finite $T_c$ while it becomes smeared ferrimagnetic one at $H>0$ in genuine ferrimagnetic region in (b).}
\label{Fig.2}
\end{figure}

We should note that on the lattices having perfectly balanced percolation cluster with $\eta_p =1$  DAFF also have a ferrimagnetic phase in its ground state. The difference with the imbalanced case is that it appears above some finite critical field ${H_{AF}}( p)$ while this field is zero if $\eta_p < 1$. The schematic ground state phase diagrams are shown in Fig. \ref{Fig.2}. The validity of these pictures follows from quite simple considerations. Let us consider the perfectly balanced AF percolation cluster. As it necessary has some imbalanced (magnetized) parts, the field will induce the energy-reducing global flipping of their spins if the magnetic moment $M$ of the part points opposite to the field and $H$ is greater than $\frac{B}{M}J$. Here $B$ is the number of AF bonds connecting the given part with the rest of percolation cluster, $J$ is AF exchange. First the large clusters with small $\frac{B}{M}$ ratio will be flipped in low fields while the field growth will induce the flipping of smaller and smaller ones. At last the remaining single spins flip along the field at $H = zJ$ ($z$ is the lattice coordination number). The corresponding jumps of sublattice magnetizations are seen in the numerical study of the ground state of $3d$ (simple cubic) DAFF with $p$ as large as 0.9 and $H > 2J$ \cite{10}.

Apparently, this process results in appearance of a nonzero magnetization of percolation cluster in fields above some $H_{AF}(p)$ and vanishing of its staggered magnetization above some greater field $H_{F}(p)$. So the ground state at ${H_{AF}}( p) < H < H_{F}(p)$ is ferrimagnetic. Yet in this case $M$ appears at $T_c$ as a secondary order parameter $M \sim L^2$ and here sharp AF transition is preserved as well as DAFF-RFIM mapping. So we may call this phase 'improper ferrimagnetic' to distinguish it from the genuine ferrimagnetic one in Fig \ref{Fig.2}b.

Still the improper ferrimagnetic ground state would cause a drastic change in the dynamics of AF phase. This is the consequence of huge degeneracy of the ferrimagnetic ground state as at rational $H/J$ there can be a huge amount of parts of the percolation cluster with $H/J=B/M$ so their flipping does not change the energy. This degeneracy is explicitly demonstrated in numerical studies of realistic DAFF systems \cite{10}, \cite{11}. At finite $T$ this results in many (nearly) degenerate minima of thermodynamic potential so the system can be trapped in each of them, depending on the previous history of $T$ and $H$ variations. The particular manifestation of these phenomena is the difference between field-cooled and zero-field-cooled thermodynamic parameters. Apparently, it would be also present in the ferrimagnetic phase of imbalanced DAFF right down to $H=0$.

Concerning the behavior of $H_{AF}( p)$ and ${H_{F}}( p)$ in Fig \ref{Fig.2} we can note that it is quite apparent that $H_{AF}(1)= H_{F}(1)= zJ$ while their diminishing to zero at $p=p_c$ in Fig \ref{Fig.2}a is the consequence of sparse structure of percolation cluster near $p_c$. Here it is divided into loosely connected parts with $B/M \to 0$ so their flipping fields also go to zero resulting in $H_{AF}( p_c)=H_{F}( p_c) = 0$. In Fig \ref{Fig.2}b ${H_{AF}}( p)$ seized to exist at $p = 1-p_c$ when, according to our surmise, $\eta_p$ becomes less than 1 in some lattices. 

The notion of ${H_{AF}}( p)$ and ${H_{F}}( p)$ behavior one can get from the results of extensive numerical studies of DAFF ground state on simple cubic and bcc lattices \cite{11}. Here the boundaries of the so called "domain state" are determined. In this state the percolation cluster of the flipped spins coexists with that of unflipped ones. Its upper boundary coincides with ${H_{F}}( p)$ while the lower one can be somewhat higher than ${H_{AF}}( p)$, yet its behavior for the simple cubic lattice \cite{11} resembles that in Fig. \ref{Fig.2}a. So, most probably, this lattice has $\eta_p =1$ for all $p$. The results for bcc lattice are less conclusive, here the percolation cluster of flipped spins can appear at rather low fields, depending on the boundary conditions and disorder realization \cite{11}. This makes bcc lattice a valid candidate for having $\eta_p <1$ (and ${H_{AF}}( p)= 0$ ) at some $p > p_c$. 

To get some notion of the DAFF thermodynamics in the ferrimagnetic phase which may result from $\eta_p < 1$ at $p<1-p_c$ we consider here the nearest-neighbor AF on the simplest hierarchical lattice, imitating the percolation cluster with fractal dimension $d = 2$ and $\eta = 1/3$. As well, it may describe qualitatively large planar aggregates of AF particles or disordered AF thin films which may have the imbalanced structure of a set of magnetic ions. The model also exhibits a number of field-induced ground state transitions marked by the magnetization jumps which are discussed above.
 
\section*{Low-field thermodynamics of hierarchical antiferromagnet}
\begin{figure}[htp]
\centering
\includegraphics[height=2.7cm]{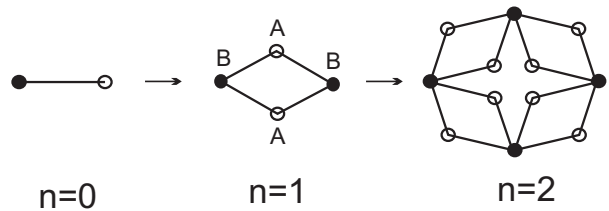}
\caption{ Construction of hierarchical lattice. It is bipartite at all levels. Different circles designate its partitioning, open circles – sublattice A, filled circles – sublattice B, $N_A \geq N_B$.}
\label{Fig.3}
\end{figure}

We consider the short-range Ising AF on the simplest "diamond" hierarchical lattice \cite{12}. Its building process is shown in Fig.\ref{Fig.3}. On the $n$-th level of hierarchy the lattice has $ N_n $ sites,
${N_n} = \frac{2}{3}\left( {{4^n} + 2} \right)$,
see Ref. \cite{13}. The coordination numbers of the sites are the powers of 2: $z = 2, 4, 8,...$. At all levels of the hierarchy the lattice is bipartite and for $n > 0$ the sites with coordination number $z = 2$ constitutes the sublattice A (open circles in Fig.\ref{Fig.3}) while the others belong to the sublattice B (filled circles), ${N_{A,n}} \geq {N_{B,n}}$. At the $n$-th level ${N_{A,n}} = 2 \cdot {4^{n - 1}}$ for $n > 0$ \cite{13}, so
${\eta _n} = ({N_n} - {N_{A,n}})/{N_{A,n}} = \left( {1 + 2 \cdot {4^{1 - n}}} \right)/3$.
We are interested in the thermodynamic limit of infinite levels of hierarchy. In this limit $\eta  = 1/3$ and fractal dimension $d = 2$ \cite{13}.

For the Ising spins $S_i = \pm 1$ placed on the sites of this lattice we consider the AF Hamiltonian 
\begin{equation}
{\mathcal{H}} = J\sum\limits_{ < i\in A, j\in B > } {{S_i}{S_j} - {H_A}\sum\limits_{i \in A} {{S_i}} }  - {H_B}\sum\limits_{j \in B} {{S_j}}                                \end{equation}
where $ < i \in A,j \in B > $ means the summation over nearest neighbors and different fields for the sublattices are introduced. This allows to find the average magnetizations of each sublattice and the order parameter for the transition. Homogeneous field corresponds to $H_A=H_B=H$. 

The usual way to get the partition function of the model is through the recursion relations for partial partition functions at different levels of hierarchy ${Z_n}(S,S')$ having fixed values of the outmost left and right spins $S$ and $S'$ \cite{12}. These relations read
\begin{equation}
{Z_{n + 1}}(S,S') = {\left[ {\sum\limits_{{S_1} =  \pm 1} {{Z_n}(S,{S_1}){e^{{h_n}{S_1}}}{Z_n}({S_1},S')} } \right]^2},                      
\end{equation} 
 ${h_n} = {H_n}/T,{\text{  }}{H_0} = {H_A}$ and${\text{  }}{H_n} = {H_B},{\text{ }}n \geq 1$.
The initial condition for them is
\begin{equation}
{Z_0}\left( {S,S'} \right) = {e^{ - KSS'}}, 
\qquad K = J/T.
\end{equation}    
Using the representation
\[
{Z_n}\left( {S,S'} \right) = \exp \frac{1}{2}\left[ {{C_n} + {u_n}SS' + \left( {{v_n} - {h_n}} \right)\left( {S + S'} \right)} \right]
\]          
we get from Eqs.(2,3) 

\begin{eqnarray}
{u_0}=-2K,\qquad v_0=h_A, \qquad C_0=0 \qquad \textit{     }  \\  \nonumber
{u_{n + 1}} = 2\ln \cosh {u_n} + \ln \left( {1 - {{\tanh }^2}{u_n}{{\tanh }^2}{v_n}} \right),  \\  \nonumber
{v_{n + 1}} = 2{v_n} + 2{\tanh ^{ - 1}}\left( {\tanh {u_n}\tanh {v_n}} \right) - 2{h_n} + {h_{n + 1}},  \\   \nonumber               
{C_{n + 1}} = 4{C_n} + {u_{n + 1}} + 4\ln \left( {2\cosh {v_n}} \right).                         
\end{eqnarray}

The last of Eqs.(4) gives for $n>0$
\[
{C_n} = {u_n} - {4^n}{u_0} + \sum\limits_{l = 0}^{n - 1} {{4^{n - l}}\left[ {{u_l} + \ln \left( {2\cosh {v_l}} \right)} \right]} 
\]
so the $n$-th level free energy per spin is
\[
  {F_n} =  - \frac{T}{{{N_n}}}\ln \sum\limits_{S,S'} {{Z_n}\left( {S,S'} \right)} {e^{{h_B}\left( {S + S'} \right)}}= \\  \nonumber
\]
\begin{equation}
-\frac{3}{4}T\sum\limits_{l = 0}^{n - 1} {{4^{ - l}}\left[ {{u_l} + \ln \left( {2\cosh {v_l}} \right)} \right]}-\frac{3}{2}J + O\left({1/{N_n}}\right)  
\end{equation}
                               
At $ H_n = 0$ the model has phase transition at $ K=K_c \approx 0.609 $ being the solution to the equation ${K_c} = \ln \cosh 2{K_c}$. ${u_c} = {2K_c}$ is the stationary point of the zero-field equations, ${u_{n + 1}} = 2\ln \cosh {u_n}$, ${v_n} = 0$. In the paramagnetic phase at $K < K_c$ ${u_n} \to 0$ for $n \to \infty $, while in the ordered phase at $K > K_c$ $ {u_n} \to \infty $.
According to above considerations the ground states of the model have magnetizations $\pm \left( {1/2} \right)$, so we may expect that the ordered phase is ferrimagnetic. To show this we consider Eqs.(4) at 
\begin{equation}
0 < ({T_c} - T)/{T_c} \equiv \tau  \ll 1, \qquad \left| {{h_n}} \right| \ll \tau.
\end{equation}
 In this case ${u_n}$ and ${v_n}$ can be found approximately in the three regions of $n$:

$
{\text{1)}}{\text{ }} 1 \leq n \leq \lambda,{u_n}-{u_c} \leq {u_c},  \left| {{v_n}} \right| \ll {\text{1}}
$
\begin{eqnarray}
{u_n} \approx {u_c} + {\kappa ^{n - 1}}\left( {{u_1} - {u_c}} \right),
\\
{v_n} \approx \frac{{{h_B} + {{\left( {2 + \kappa } \right)}^{n - 1}}\kappa \tilde h}}{1 + \kappa},\qquad \tilde h = {h_B} - \left( {\kappa  + 1} \right){h_A}                                       
\end{eqnarray}
\[
 {u_1}= 2\ln \cosh 2K, \qquad \kappa = 2\tanh 2{K_c} \approx 1.68,
\]
The value of $\lambda $ is defined by
\begin{equation}
{u_\lambda } = 2{u_c},\qquad \kappa ^{ - \lambda } = \frac{{{u_1} - {u_c}}}{{\kappa {u_c}}} \approx \tau 
\end{equation}                                                                           
while $\vert {{v_n}} \vert \ll {\text{1}}$ requires 
\begin{equation}
\vert{\tilde h}\vert{\left( {2 + \kappa } \right)^\lambda } = \vert{\tilde h}\vert{\tau ^{ - \ln \left( {2 + \kappa } \right)/\ln \kappa }} \ll 1.
\end{equation}

$ 
{\text{2)}}{\text{ }} \lambda < n \leq \mu, 1\ll u_n \ll \vert v_n \vert
$
\begin{eqnarray}
{u_n} \approx {2^{n - \lambda  + 1}}{u_c} - \sum\limits_{k = \lambda }^{n - 1}{{2^{n - k}}\ln 2\cosh {v_k}}, 
\\ 
{v_n} \approx \frac{{{h_B}}}{3} + {4^{n - \lambda }}\left( {{v_\lambda } - \frac{{{h_B}}}{3}} \right),
\\ 
{v_\lambda } \approx \frac{{{{\left( {2 + \kappa } \right)}^{\lambda  - 1}}\kappa \tilde h}}{{\left( {1 + \kappa } \right)}}=\frac{{\kappa \tilde h}}{{\left( {1 + \kappa } \right)\left( {2 + \kappa } \right){\tau ^{\ln \left( {2 + \kappa } \right)/\ln \kappa }}}}                   
\end{eqnarray}
The value of $\mu$ is defined by the equation $u_\mu =\vert v_\mu \vert$. As
${u_\mu } \approx {2^{\mu  - \lambda  + 1}}{u_c} - \sum\limits_{k = \lambda }^{\mu  - 1} {{2^{\mu  - k}}\ln 2\cosh {4^{k - \lambda }}{v_\lambda }}  \approx {2^{\mu  - \lambda }}2\left( {{u_c} - \ln 2} \right) + {4^{\mu  - \lambda }}{v_\lambda },$  ${v_\mu } \approx {4^{\mu  - \lambda }}{v_\lambda }$,
we get
\begin{equation}
{2^\mu } = \frac{{\left( {{u_c} - \ln 2} \right){2^\lambda }}}{{\left| {{v_\lambda }} \right|}},{\text{    }}{u_\mu } = \left| {{v_\mu }} \right| = \frac{{{{\left( {{u_c} - \ln 2} \right)}^2}}}{{\left| {{v_\lambda }} \right|}}.                                                 
\end{equation}
\begin{equation}
{\text{3)}}{\text{ }}  \mu > n, \vert {{v_n}} \vert \gg {\text{1}}, {u_n} \approx 0, {v_n} \approx 2^{n - \mu}v_\mu .                                                         
\end{equation}

Note that in the sums we consider the large numbers $\lambda $ and $\mu $ as integers neglecting its fractional parts. 

Using the above approximations for ${u_n}$ and ${v_n}$ we can find from Eq.(5) free energy in the thermodynamic limit near the transition point in a small field (cf. Eqs.(6), (10)). Thus, dividing the sum in (5) in three parts and $n=0$ term, 
\[ 
- \frac{4}{3}\frac{F}{T} = \ln 2\cosh {h_A} + {\Sigma _\lambda } + {\Sigma _{\lambda \mu }} + {\Sigma _\mu },
\]
we get
\begin{eqnarray*}
{\Sigma _\lambda } = \sum\limits_{n = 1}^\lambda  {\left[ {{4^{ - n}}\left( {{u_c} + \ln 2} \right) + {u_c}\tau {\left( \frac{\kappa }{4} \right)}^n}\right]}+ \\
+\sum\limits_{n = 1}^\lambda {\frac{\tilde h^2}{2{\left( 1 + \kappa \right)}^2 {\left( 2 + \kappa \right)}^2}{\left( 1 + \frac{\kappa }{2} \right)}^{2n}}      \\
   \approx \frac{1}{3}\left( \ln 2 + {u_c} \right) + \frac{\kappa }{4 - \kappa }{u_c}\tau  - \frac{{{4^{ - \lambda }}}}{3}\left( {\frac{{2 + \kappa }}{{4 - \kappa }}2{u_c} + \ln 2} \right) \\
   + \frac{{\kappa {{\tilde h}^2}}}{{2\left( {4 + \kappa } \right){{\left( {1 + \kappa } \right)}^2}}}{\left( {1 + \frac{\kappa }{2}} \right)^{2\lambda }}
\end{eqnarray*}

\begin{eqnarray*}
  {\Sigma _{\lambda \mu }} = \sum\limits_{n = \lambda  + 1}^\mu  {\left( {{2^{ - n - \lambda  + 1}}{u_c} + {4^{ - n}}\ln 2\cosh {v_n}} \right)}- \\
- \sum\limits_{n = \lambda  + 1}^\mu  {{4^{ - n}}\sum\limits_{k = \lambda }^{n - 1} {{2^{n - k}}\ln 2\cosh {v_k}} }  \hfill \\
   = \sum\limits_{n = \lambda  + 1}^\mu  {\left( {{2^{ - n - \lambda  + 1}}{u_c} + {4^{ - n}}\ln 2\cosh {v_n}} \right)}-  \\
- \sum\limits_{k = \lambda }^\mu  {\left( {{4^{ - k}} - {2^{ - k - \mu }}} \right)\ln 2\cosh {v_k}}  \hfill \\
   = {4^{ - \lambda }}2{u_c} - {4^{ - \mu }}{u_\mu } + {4^{ - \mu }}\ln 2\cosh {v_\mu } - {4^{ - \lambda }}\ln 2\cosh {v_\lambda } \\
 \approx {4^{ - \lambda }}\left( {2{u_c} - \ln 2} \right) \hfill \\ 
\end{eqnarray*} 

${\Sigma _\mu } \approx \sum\limits_{n = \mu  + 1}^\infty  {{4^{ - n}}\left| {{v_n}} \right|}  \approx {4^{ - \mu }}\left| {{v_\mu }} \right| = {4^{ - \lambda }}\left| {{v_\lambda }} \right|$.

Here we used $\left| {{v_\lambda }} \right| \ll 1,{\text{ }}\left| {{v_\mu }} \right| \gg 1$, Eqs.(7, 8, 11, 12, 15) and the relation following from Eq.(11),
\[\sum\limits_{k = \lambda }^{\mu  - 1} {{2^{ - k - \mu }}\ln 2\cosh {v_k}}  = {2^{ - \lambda  - \mu  + 1}}{u_c} - {4^{ - \mu }}{u_\mu }.\]
Finally we have from Eqs.(5, 9, 13, 14)
\begin{equation}
F/{T_c} \approx  - {K_c}/2 - \ln 2 + \tau {s_c} - a{\tau ^{2 - \alpha }} - b{\tau ^\beta }\left| {\tilde h} \right| - c{\tau ^{ - \gamma }}{\tilde h^2},                                \end{equation}      

${s_c} = \ln 2 - 2{K_c}\frac{{\kappa  - 1}}{{4 - \kappa }} \approx 0.34$, $a = 2{K_c}\frac{{5 - 2\kappa }}{{4 - \kappa }} - \ln 2 \approx 0.17$, $b = \frac{{3\kappa }}{{4\left( {2 + \kappa } \right)\left( {1 + \kappa } \right)}} \approx 0.13$,
$c = \frac{{3\kappa }}{{8\left( {4 + \kappa } \right){{\left( {1 + \kappa } \right)}^2}}} \approx 0.015$.

\begin{eqnarray}
\alpha  = 2 - \frac{{\ln 4}}{{\ln \kappa }} \approx  - 0.67, \\ \nonumber
 \beta  = \frac{{\ln 4 - \ln \left( {2 + \kappa } \right)}}{{\ln \kappa }} \approx 0.16, \\ \nonumber
 \gamma  = \frac{{2\ln \left( {2 + \kappa } \right) - \ln 4}}{{\ln \kappa }} \approx 2.35
\end{eqnarray}
 In homogeneous field $\tilde h =  - \kappa h$ (cf. Eq.(8)) so $F$ in Eq.(16) has the standard scaling form of a ferromagnet with spontaneous magnetization $m \sim {\tau ^\beta }$ and divergent susceptibility $\chi  \sim {\tau ^{ - \gamma }}$. This expression is valid at $0 < \tau  \ll 1$, $\left| h \right| \ll {\tau ^{\beta  + \gamma }}$, cf. Eq.(10). Scaling indices (17) obey the usual relation $\alpha  + 2\beta  + \gamma  = 2$. 
Negative $ \alpha $ means that specific heat is finite at the transition point and has a cusp at $T_c$ . Note also that ${s_c}$ is the entropy at the transition point. So this AF system looks like genuine ferromagnet, even featuring the absence (smearing) of transition in a finite field. The last is evident as the nontrivial stationary point of finite-field recursion relations (4) cannot be reached from any initial conditions. 

Yet the dependence of $F$ on $\tilde h$ from Eq.(8) shows that true order parameter for the transition is a linear combination of ${M_A} = \sum\limits_{i \in A} {{S_i}} $ and ${M_B} = \sum\limits_{i \in B} {{S_i}} $ conjugate with $\tilde H = {H_B} - \left( {\kappa  + 1} \right){H_A}$. To distinguish the order parameter in the Hamiltonian (1) we perform a coordinate rotation in $2d$ space of vectors ${\mathbf{H}} = \left( {{H_A},{H_B}} \right)$ and ${\mathbf{M}} = \left( {{M_A},{M_B}} \right)$ to bring the term $ - {\mathbf{MH}}$ in (1) to the form $ - {\mathbf{MH}} =  - \tilde M\tilde H - M'H'$ where
\begin{eqnarray*}
\tilde M = \frac{{{M_B} - \left( {\kappa  + 1} \right){M_A}}}{{1 + {{\left( {\kappa  + 1} \right)}^2}}}, M' = \frac{{\left( {\kappa  + 1} \right){M_B} + {M_A}}}{{1 + {{\left( {\kappa  + 1} \right)}^2}}}, \\
  H' = \left( {\kappa  + 1} \right){H_B} + {H_A}
\end{eqnarray*}
Thus $\tilde M$ is the order parameter while $M'$ and $H'$ are non-critical variables. Hence, $\left\langle {M'} \right\rangle  = 0$ at $\bm H \to 0$ so the spontaneous magnetic moments of the sublattices obey the relation $\left\langle {{M_A}} \right\rangle  =  - \left( {\kappa  + 1} \right)\left\langle {{M_B}} \right\rangle$. Then for the spontaneous magnetizations ${m_\nu } = \left\langle {{M_\nu }} \right\rangle /{N_\nu }$ ($\nu  = {\text{ }}A,{\text{ }}B$) we have 
\[
{m_A} =  - \eta \left( {\kappa  + 1} \right){m_B} \approx  - 0.9{m_B}
\]
This differs from the ground state relation ${m_A} =  - {m_B}$. We may suggest that this is a consequence of critical fluctuations diminishing ${m_A}$ more strongly than ${m_B}$ as all sites of sublattice A have the lowest coordination number ${z_A} = {\text{2}}$. To some extent this effect would be present in all dilute AF on imbalanced bipartite graphs since the sublattice A with larger amount of spins would necessary have lower average coordination number ${\bar z_A} = C/{N_A} < {\bar z_B} = C/{N_B}$. Here $C$ is the number of bonds and we used the fact that all bonds are of A-B type. Thus ${\bar z_A} = \eta {\bar z_B}$ so the lower $\eta$ the stronger can be the fluctuation-induced disbalance between ${m_A}$ and ${m_B}$ near $T_c$. At $\eta =1$ this effect vanishes so its observation in neutron-diffraction experiments can certify the onset of imbalance in the magnetic percolation cluster.
\section*{Ground state transitions}
Here we assume $H_A=H_B=H$.  At $T = 0$ we define
${\tilde u_n} = \mathop {\lim }\limits_{T \to 0} T{u_n},{\text{   }}{\tilde v_n} = \mathop {\lim }\limits_{T \to 0} T{v_n},{\text{   }}E = \mathop {\lim }\limits_{T \to 0} F$
to obtain from Eqs.(4, 5)
\begin{eqnarray}
\tilde u_0 =  - 2J \qquad         \tilde v_0 = H \nonumber \\ 
{\tilde u}_{n + 1} = 2\left( {\left| {\tilde u}_n \right| - \left| {\tilde v}_n \right|} \right)\vartheta \left( {\left| {\tilde u}_n \right| - \left| {\tilde v}_n \right|} \right),   \qquad  \\
{\tilde v}_{n + 1} = 2{\tilde v}_n + 2\min \left( \left| {\tilde u}_n \right|,\left| {\tilde v}_n \right| \right)\operatorname{sgn} \left( {\tilde u}_n{\tilde v}_n \right) - H,   
\end{eqnarray}                         
\[
- (4/3)E = H + \sum_{n = 1}^{\infty}  {{4^{ - n}}\left( {{{\tilde u}_n} + \left| {{{\tilde v}_n}} \right|} \right)} 
\]                                 
$\vartheta $ in Eq.(18) is Heaviside's step function. Solving Eqs.(18, 19), we get
\begin{eqnarray*}
- (4/3)E = H + \sum_{n = 1}^{\infty}  {{4^{ - n}}\left| {H - {2^{n + 1}}J} \right|},     
\\
m =  - \frac{{\partial E}}{{\partial H}} = \frac{3}{4}\left[ {1 + \sum_{n = 1}^{\infty}   {{4^{ - n}}\operatorname{sgn} \left( {H - {2^{n + 1}}J} \right)} } \right].
\end{eqnarray*}  
So at $H \ne {2^k}J$    
\[m = {2^{ - 1}}\vartheta \left( {2J - H} \right) + \left( {1 - 2 \cdot {4^{ - r}}} \right)\vartheta \left( {H - 2J} \right),
\]
where $r = \left[ {{{\log }_2}\left( {H/J} \right)} \right]$ is an integer part of ${\log _2}\left( {H/J} \right)$. At ${H_r} = {2^r}J$,  $r \geqslant 2$, we have ${m_r} = 1 - 6 \cdot {4^{ - r}}$. 
\begin{figure}[htp]
\centering
\includegraphics[height=5cm]{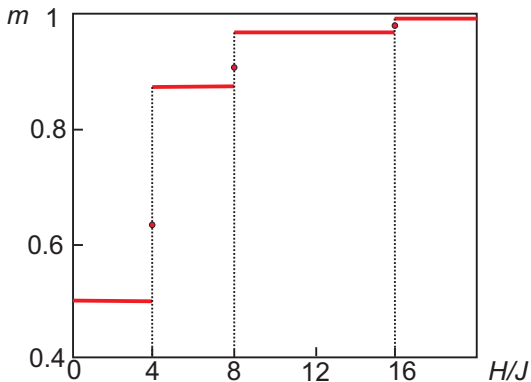}
\caption{Field dependence of the ground state magnetization.} \label{Fig.4}
\end{figure}
Field dependence of the ground state magnetization is shown in Fig.\ref{Fig.4}. Due to the imbalance ratio $\eta  = 1/3$ the system has spontaneous magnetization $m = 1/2$ at $H \to +0$. Unfortunately, the data on the percolation cluster magnetization on cubic and bcc lattices are totally absent in Ref. \cite{11} which deprives us of the opportunity to decide if there is the imbalance in real $3d$ percolation clusters at $p_c < p < 1 - p_c$. 

 The jumps at ${H_r} = {2^r}J$ result from the flipping along the field of single spins in sublattice B having the coordination number $2^r$. As we discussed above in dilute crystalline lattices there are many more jumps appearing at rational values of $H/J$ where flipping of the magnetized parts of percolation cluster takes place \cite{10}. Such jumps were observed in low-$T$ experiments in $Fe_pZn_{1-p}F_2$ \cite{14}.  
\section*{Discussion and conclusions}
The decades of experimental investigations of dilute Ising AF have shown that   DAFF - RFIM correspondence works reasonably well at low dilution and low fields \cite{15}. Meanwhile the field-induced rounding of the transition appears at lower $p$ which is impossible in the case of the AF ordered phase. One explanation assumes that this is nonequilibrium effect due to the pinning of AF domain walls by the vacancies which results in very slow relaxation to the equilibrium AF structure \cite{16}. Also one may suggest that AF transition transforms at lower $p$ into a spin-glass one \cite{15}, \cite{17}. 

Here we argue that one more reason for the vanishing of AF transition could be the imbalance of percolation cluster which makes transition ferrimagnetic at $H=0$ and smeared at finite fields. Just these phenomena were found in $Fe_pZn_{1-p}F_2$ \cite{5} - \cite{7} and several other dilute AFs \cite{8}, \cite{9}. Also the inspection of neutron-diffraction data on metastability and domain formation in $Fe_pZn_{1-p}F_2$ family \cite{18} makes authors to conclude that AF order vanishes right at $p = 1- p_c$. AF region in the $H-p$ phase diagram of these compounds in Ref. \cite{18} is quite similar to that in Fig.\ref{Fig.2}b. Thus our surmise of possible imbalance of percolation cluster at $p < 1- p_c$ seems to be true for bcc lattice. 

This implies that qualitative features of the considered here model could apply to the thermodynamics of $Fe_pZn_{1-p}F_2$ compounds with vacancies' percolation. They are a small scaling index of remanent magnetization, rather high index $\gamma $, large negative $\alpha$ and disbalance in the sublattice magnetizations near $T_c$.  But to observe these features of ferrimagnetic transition the measurements in ultra-low fields (same as in Refs. \cite{5}-\cite{9}) are needed to avoid its smearing. Also the irreversibility should be taken into account as the theoretical results refer only to the most stable state, which seemingly is a field-cooled one in the ferrimagnetic phase. 
 
Now we do not know on which lattices DAFF would also have the phase diagram of Fig.\ref{Fig.2}b. Moreover, the exact form of ${H_{AF}}( p)$ and ${H_{F}}( p)$ is not known for the variety of crystalline lattices of the known easy-axes antiferromagnets. Yet investigation of DAFF ground state in Ref. \cite{11} shows that their determination is feasible with modern numerical methods. Such studies and further experiments revealing the details of low-$T$ and low-$H$ behavior of magnetization may help to elucidate the nature of transition in nearest-neighbor dilute AF.

Author gratefully acknowledges useful discussions with V.P. Sakhnenko and M.P. Ivliev.

\end{document}